\documentclass[aps,pra,showpacs,floatfix]{revtex4}
\usepackage{graphicx}
\usepackage{subfigure}
\begin{document}

\title{Phase Diffusion in Quantum Dissipative 
Systems} 
\author{Subhashish Banerjee}
\email{subhashishb@rri.res.in}
\affiliation{Raman Research Institute, Sadashiva Nagar,
Bangalore - 560 080, India}
\author{R. Srikanth}
\email{srik@rri.res.in}
\affiliation{Poornaprajna Institute of Scientific Research,
Devanahalli, Bangalore- 562 110, India}
% Sadashiva Nagar, 
\affiliation{Raman Research Institute, Sadashiva Nagar,
Bangalore - 560 080, India}

%%\date{21 November 2006}

\begin{abstract}
We  study the dynamics  of the  quantum phase  distribution associated
with the reduced density matrix of a system for a number of situations
of practical importance, as the  system evolves under the influence of
its  environment,  interacting  via  a quantum  nondemoliton  type  of
coupling, such that there  is decoherence without dissipation, as well
as  when it  interacts  via a  dissipative  interaction, resulting  in
decoherence as well as dissipation. The system is taken to be either a
two-level atom  (or equivalently, a  spin-$1/2$ system) or  a harmonic
oscillator,  and the  environment is  modeled  as a  bath of  harmonic
oscillators, starting out in a  squeezed thermal state.  The impact of
the different environmental parameters  on the dynamics of the quantum
phase  distribution for  the system  starting out  in  various initial
states,  is  explicitly  brought  out.  An  interesting  feature  that
emerges from our  work is that the relationship  between squeezing and
temperature effects  depends on  the type of  system-bath interaction.
In the  case of quantum  nondemolition type of  interaction, squeezing
and temperature  work in tandem,  producing a diffusive effect  on the
phase   distribution.   In   contrast,  in   case  of   a  dissipative
interaction,  the  influence of  temperature  can  be counteracted  by
squeezing, which manifests as  a resistence to randomization of phase.
We  make use  of the  phase  distributions to  bring out  a notion  of
complementarity in  atomic systems.  We  also study the  dispersion of
the  phase using  the  phase distributions  conditioned on  particular
initial states of the system.
\end{abstract} 

\pacs{03.65.Yz, 42.50.Ct} 

\maketitle

\section{Introduction}

Open quantum systems  are ubiquitous in the sense  that any system can
be thought  of as  being surrounded by  its environment  (reservoir or
bath) which influences its dynamics.  They provide a natural route for
discussing damping and dephasing. One of the first testing grounds for
open system  ideas was in quantum optics  \cite{wl73}. Its application
to other areas gained momentum  from the works of Caldeira and Leggett
\cite{cl83}, and Zurek \cite{wz93},  among others.  Depending upon the
system-reservoir  ($S-R$)  interaction, open  systems  can be  broadly
classified into two categories,  viz., quantum non-demolition (QND) or
dissipative.  A  particular type of quantum  nondemolition (QND) $S-R$
interaction is  given by a class of  energy-preserving measurements in
which  dephasing  occurs without  damping  the  system.   This may  be
achieved when  the Hamiltonian $H_S$  of the system commutes  with the
Hamiltonian  $H_{SR}$  describing  the  system-reservoir  interaction,
i.e.,  $H_{SR}$  is  a  constant  of the  motion  generated  by  $H_S$
\cite{sgc96, mp98,  gkd01}.  A dissipative  open system would  be when
$H_S$ and  $H_{SR}$ do not  commute resulting in dephasing  along with
damping \cite{bp02}.  A prototype of dissipative open quantum systems,
having many  applications, is the quantum Brownian  motion of harmonic
oscillators.   This   model  was  studied  by   Caldeira  and  Leggett
\cite{cl83} for  the case  where the system  and its  environment were
initially  separable.  The  above  treatment of  the quantum  Brownian
motion was generalized to  the physically reasonable initial condition
of  a mixed  state of  the  system and  its environment  by Hakim  and
Ambegaokar  \cite{ha85},  Smith  and  Caldeira  \cite{sc87},  Grabert,
Schramm and  Ingold \cite{gsi88}, and  for the case  of a system  in a
Stern-Gerlach potential \cite{sb00}, and also for the quantum Brownian
motion  with  nonlinear  system-environment  couplings  \cite{sb03-2},
among others.

The  interest  in  the  relevance  of open  system  ideas  to  quantum
information has  increased in recent  times because of  the impressive
progress made on the experimental front in the manipulation of quantum
states of  matter towards  quantum information processing  and quantum
communication.  Myatt {\em et al.} \cite{myatt} and Turchette
{\em et al.} \cite{turch}  have performed a series of experiments
in  which they  induced decoherence  and  decay by  coupling the  atom
(their system-$S$) to engineered reservoirs, in which the coupling to,
and the  state of, the  environment are controllable. An
experiment  reported in Ref.  \cite{jb03} demonstrated  and completely
characterized a  QND scheme for making  a nondeterministic measurement
of  a single  photon  nondestructively using  only  linear optics  and
photo-detection of ancillary modes, to induce a strong nonlinearity at
the single  photon level.  The  dynamics of decoherence  in continuous
atom-optical QND  measurements has been  studied by Onofrio  and Viola
\cite{vo98}.    In  addition  to   its  relevance   in  ultrasensitive
measurements,  a  QND  scheme   provides  a  way  to  prepare  quantum
mechanical states which may otherwise  be difficult to create, such as
Fock states  with a specific number  of particles.  It  has been shown
that the  accuracy of atomic  interferometry can be improved  by using
QND  measurements of  the  atomic  populations at  the  inputs to  the
interferometer \cite{kbm98}.  QND systems  have also been proposed for
engineering quantum dynamical evolution of a system with the help of a
quantum  meter \cite{ca05}.   In a  recent  study of  QND open  system
Hamiltonians for  two different models of  the environment describable
as baths of either oscillators or spins, an interesting connection was
found  between the  energy-preserving QND  Hamiltonians and  the phase
space area-preserving canonical transformations \cite{sb07}.

A class of observables that  may be measured repeatedly with arbitrary
precision,  with the  influence of  the measurement  apparatus  on the
system being confined strictly to the conjugate observables, is called
QND or back-action evasive observables \cite{bvt80, bk92, wm94, zu84}.
Such a measurement scheme was  originally introduced in the context of
the detection  of gravitational  waves \cite{ct80, bo96}.   The energy
preserving measurements, referred to above, form an important class of
such a  general QND measurement scheme. Since  they describe dephasing
without dissipation, a study of phase diffusion in such a situation is
important from the context of a number of experimental situations.
 
The  quantum description  of  phases \cite{sch93,pp98}  has  a long  history
\cite{pad27, sg64,  cn68, pb89, ssw90}. Pegg  and Barnett \cite{pb89},
following Dirac \cite{pad27}, carried out a polar decomposition of the
annihilation  operator and  defined a  hermitian phase  operator  in a
finite-dimensional  Hilbert space.  In  their scheme,  the expectation
value of  a function of the phase  operator is first carried  out in a
finite-dimensional Hilbert  space, and then the dimension  is taken to
the limit of  infinity. However, it is not  possible to interpret this
expectation value as that of  a function of a hermitian phase operator
in  an  infinite-dimensional  Hilbert  space \cite{ssw91,  mh91}.   To
circumvent  this problem, the  concept of  phase distribution  for the
quantum phase has been  introduced \cite{ssw91,as92}.  In this scheme,
one associates  a phase  distribution to a  given state such  that the
average of  a function  of the phase  operator in the  state, computed
with  the  phase distribution,  reproduces  the  results  of Pegg  and
Barnett.

A study of the quantum phase  diffusion in a number of QND systems was
carried out in Ref. \cite{sb06} using the phase distribution approach.
In  this work  we extend  the  above study  to include  the effect  of
dissipation on  phase diffusion.  Throughout  this paper, the  bath is
assumed to  be a  collection of harmonic  oscillators starting  from a
squeezed  thermal initial  state.  An  advantage of  using  a squeezed
thermal  bath is  that the  decay rate  of quantum  coherences  can be
suppressed   leading   to   preservation  of   non-classical   effects
\cite{kw88, kb93,bg06}. It has also been shown to modify the evolution
of the geometric phase of two-level atomic systems \cite{bsri06}.  The
plan of  the paper is  as follows.  In  Section II, we  recollect some
results  on  the  quantum  phase  distribution  in  QND  systems  from
\cite{sb06, bg06}.   We extend the previous expressions,  for a single
two-level  atomic system,  to  the  case of  two  two-level atoms  and
further plot  the quantum phase distribution for  ten two-level atoms.
Following Agarwal  and Singh \cite{as96} we also  introduce the number
distribution  and use  it  to discuss  the  complementary between  the
number and phase  distributions. In Section III, we  study the quantum
phase distribution  of a two-level atomic system  interacting with its
bath via  a dissipative interaction.   The evolution is governed  by a
Lindblad equation.   The phase distribution is studied  for the system
initially  (a)  in an  atomic  coherent state  and  (b)  in an  atomic
squeezed  state.   For  the   system  in  an  atomic  coherent  state,
complementarity  between   the  number  and   phase  distributions  is
discussed. In Section IV, the quantum phase distribution of the system
of a harmonic oscillator, in  a dissipative interaction with its bath,
is obtained. In Section V, an application is made of the quantum phase
distributions  obtained for  various initial  system states  and $S-R$
interactions, to study the corresponding phase dispersion.  In Section
VI, we present our conclusions.

\section{Quantum Phase Distribution: QND}

Here we recapulate, from \cite{sb06}, the results of Quantum Phase 
Distributions for a two-level atomic system as well as that of a
harmonic oscillator which undergo interaction with their environments
via a QND type of interaction. We consider the following 
Hamiltonian which models the 
interaction of a system with its environment, modeled as a bath 
of harmonic oscillators, via a QND type of coupling 
\cite{bg06}:
\begin{eqnarray}
H & = & H_S + H_R + H_{SR} \nonumber\\ & = & H_S + 
\sum\limits_k \hbar \omega_k b^{\dagger}_k b_k + H_S 
\sum\limits_k g_k (b_k+b^{\dagger}_k) + H^2_S \sum\limits_k 
{g^2_k \over \hbar \omega_k}. \label{2.1} 
\end{eqnarray} 
Here $H_S$, $H_R$ and $H_{SR}$ stand for the Hamiltonians of 
the system, reservoir and system-reservoir interaction, 
respectively. The $g_k$'s are dimensionless coupling constants.
The last term on the right-hand side of Eq. (1) 
is a renormalization inducing `counter term'. Since $[H_S, 
H_{SR}]=0$, (1) is of QND type. Here $H_S$ is a generic system 
Hamiltonian which will be used subsequently to 
model different physical situations. Assuming separable initial
conditions with the bath being initially in a squeezed thermal state
and tracing over the bath degrees of freedom, the reduced density matrix
of the system of interest $S$, in the system eigenbasis, is obtained as 
\cite{bg06}
\begin{eqnarray}
\rho^s_{nm} (t) & = & e^{-{i \over \hbar}(E_n-E_m)t} e^{
i(E^2_n-E^2_m)\eta(t)}
\times \exp \Big[-(E_m-E_n)^2 \gamma(t) \Big] 
\rho^s_{nm} (0), \label{2.2} 
\end{eqnarray}
where 
\begin{equation}
\eta (t) = - \sum\limits_k {g^2_k \over \hbar^2 \omega^2_k} 
\sin (\omega_k t), \label{2.3} 
\end{equation}
and
\begin{equation}
\gamma (t) = {1 \over 2} \sum\limits_k {g^2_k \over \hbar^2 
\omega^2_k} \coth \left( {\beta \hbar \omega_k \over 2} \right) 
\left| (e^{i\omega_k t} - 1) \cosh (r_k) + (e^{-i\omega_k t} - 
1) \sinh (r_k) e^{i2\Phi_k} \right|^2. \label{2.4} 
\end{equation}
For the reservoir $R$ to be considered as a proper bath causing
decoherence and (possibly)
dissipation, we need to assume a `quasi-continuous' bath 
spectrum with spectral density $I(\omega)$ such that 
for an arbitrary function $f(\omega)$ the continuum limit implies
\cite{gkd01}
\begin{equation}
\sum\limits_k {g^2_k \over \hbar^2} f(\omega_k) \longrightarrow 
\int\limits^{\infty}_0 d\omega I(\omega) f(\omega). \label{continuum} 
\end{equation}
We consider the case of an Ohmic bath with spectral density
\begin{equation}
I(\omega) = {\gamma_0 \over \pi} \omega e^{-\omega/\omega_c}, 
\label{2.5} 
\end{equation}
where  $\gamma_0$,   having  the  dimension   of  $1/({\rm energy})^2$
\cite{gkd01},  and
$\omega_c$ are  two bath parameters characterizing  the quantum noise.
Using Eqs. (\ref{continuum}) and (\ref{2.5}) in Eq. (\ref{2.3}),
we obtain \cite{bg06}
\begin{equation}
\eta (t) = -{\gamma_0 \over \pi} \tan^{-1} (\omega_c t). 
\label{2.6} 
\end{equation}
Using Eqs. (\ref{continuum}), (\ref{2.5}) in Eq. (\ref{2.4}) and using
the $T = 0$ limit, $\gamma (t)$ is obtained as \cite{bg06}
\begin{eqnarray}
\gamma (t) & = & {\gamma_0 \over 2\pi} \cosh (2r) \ln 
(1+\omega^2_c t^2) - {\gamma_0 \over 4\pi} \sinh (2r) \ln 
\left[ {\left( 1+4\omega^2_c(t-a)^2\right) \over \left( 1+ 
\omega^2_c (t-2a)^2 \right)^2} \right] \nonumber \\ & & - 
{\gamma_0 \over 4\pi} \sinh (2r) \ln (1+4a^2\omega^2_c) , 
\label{2.7} 
\end{eqnarray}
where  the resulting integrals are defined only for
$t  >  2a$ \cite{grad}.   
Using  Eqs.   (\ref{continuum}),  (\ref{2.5})  in
Eq. (\ref{2.4}) and using the high $T$ limit, $\gamma (t)$ is obtained
as \cite{bg06}
\begin{eqnarray} 
\gamma (t) & = & {\gamma_0 k_BT \over \pi \hbar \omega_c} \cosh 
(2r) \left[ 2\omega_c t \tan^{-1} (\omega_c t) + \ln \left( {1 
\over 1+\omega^2_c t^2} \right) \right] \nonumber \\ & & - 
{\gamma_0 k_BT \over 2\pi \hbar \omega_c} \sinh (2r) \Bigg[ 
4\omega_c (t-a) \tan^{-1} \left( 2\omega_c (t-a) \right) 
\nonumber \\ & & - 4\omega_c (t-2a) \tan^{-1} \left( \omega_c 
(t-2a) \right) + 4a\omega_c \tan^{-1} \left( 2a\omega_c \right) 
\nonumber \\ & & + \ln \left( {\left[ 1+\omega^2_c (t-2a)^2 
\right]^2 \over \left[ 1+4\omega^2_c (t-a)^2 \right]} \right) + 
\ln \left( {1 \over 1+4a^2\omega^2_c} \right) \Bigg] , 
\label{eq:gamma} 
\end{eqnarray} 
where, again, the resulting integrals are defined
for $t  > 2a$ \cite{grad}. Here we have for  simplicity taken the squeezed
bath parameters as
\begin{eqnarray} 
\cosh \left( 2r(\omega) \right) & = & \cosh (2r),~~ \sinh 
\left( 2r (\omega) \right) = \sinh (2r), \nonumber\\ \Phi 
(\omega) & = & a\omega, \label{eq:a} 
\end{eqnarray} 
where $a$ is  a constant depending upon the  squeezed bath.  Note that
the  results pertaining to  a thermal  bath can  be obtained  from the
above equations by setting the  squeezing parameters $r$ and $\Phi$ to
zero.  It  is interesting  to  note that  in  the  context of  quantum
information, the  open system effect  depicted in this Section  can be
modeled by a  familiar quantum noisy channel, viz.,  the phase damping
channel \cite{bsri06, deleter, nc00}.

\subsection{Two-Level Atomic Systems}

Here we  consider the case where  our system $S$ is  a two-level atom.
The system Hamiltonian $H_S$ is
\begin{equation}
H_S = {\hbar \omega \over 2} \sigma_z, \label{2a.1}
\end{equation}
where $\sigma_z$ is the  usual Pauli matrix.
The form of  the system
Hamiltonian   $H_S$,    Eq.   (\ref{2a.1}),   when    substituted   in
Eq. (\ref{2.1})  has been used  in the context of  quantum computation
\cite{wu95,  ps96, dd95}. In the context of a system of multiple two-level
atoms, which is equivalent to an angular momentum system, we set
$H_S = \hbar \omega J_z$.  The  Wigner-Dicke state  \cite{rd54, jr71,
at72} $|j, m  \rangle$, which are the simultaneous  eigenstates of the
angular momentum operators $J^2$ and  $J_z$, serve as the basis states
for $H_S$, and we have
\begin{eqnarray}
H_S|j, m \rangle & = & \hbar \omega m |j, m \rangle \nonumber 
\\ & = & E_{j,m} |j, m \rangle. \label{2a.2} 
\end{eqnarray} 
Here $-j \leq m \leq j$. Using this basis and the above 
equation in Eq. (\ref{2.2}) we obtain the reduced density matrix 
of the system as 
\begin{equation}
\rho^s_{jm,jn}(t) = e^{-i \omega (m-n)t} e^{i(\hbar \omega 
)^2 (m^2 - n^2) \eta(t)} e^{-(\hbar 
\omega )^2 (m - n)^2 \gamma(t)} \rho^s_{jm,jn}(0). \label{2a.3} 
\end{equation}
Following Agarwal and Singh \cite{as96} we introduce the phase 
distribution ${\cal P}(\phi)$, $\phi$ being related to the 
phase of the dipole moment of the system, as 
\begin{equation} 
{\cal P}(\phi) = {2j+1 \over 4 \pi} \int_{0}^{\pi} d\theta 
\sin(\theta) Q(\theta, \phi), \label{2a.4} 
\end{equation}
where ${\cal P}(\phi)> 0$ and is normalized to unity, i.e., 
$\int_{0}^{2\pi} d\phi {\cal P}(\phi) = 1$. Here $Q(\theta, 
\phi)$ is defined as 
\begin{equation}
Q(\theta, \phi) = \langle \theta, \phi|\rho^s| \theta, \phi 
\rangle, \label{2a.5} 
\end{equation}
where $|\theta, \phi \rangle$ are the atomic coherent states 
\cite{mr78, ap90} given by an expansion over the Wigner-Dicke 
states \cite{at72} as 
\begin{equation}
|\theta, \phi \rangle = \sum\limits_{m= -j}^j \left(\matrix{2j 
\cr j + m}\right)^{1 \over 2} (\sin(\theta / 2))^{j+m} 
(\cos(\theta / 2))^{j-m} |j, m \rangle e^{-i(j + m) \phi}. 
\label{2a.6} 
\end{equation} 
Using Eq. (\ref{2a.5}) in Eq. (\ref{2a.4}), with insertions of 
partitions of unity in terms of the Wigner-Dicke states, we can 
write the phase distribution function as 
\begin{eqnarray} 
{\cal P}(\phi) &=&  {2j+1 \over 4 \pi} \int_{0}^{\pi} d\theta 
\sin \theta \sum\limits_{n,m= -j}^{j} \langle \theta, \phi |j, 
n \rangle \nonumber\\ & & \times \langle j, n| \rho^s (t)| j, m 
\rangle \langle j, m| \theta, \phi \rangle. \label{2a.7} 
\end{eqnarray} 
Now we take up two physically interesting initial conditions for the
system $S$.

\subsubsection{System initially in an atomic coherent state}

Here we consider the system $S$ to be initially in an atomic 
coherent state which is the atomic analogue of the Glauber 
coherent state \cite{at72}. Thus the initial system density 
matrix is 
\begin{equation}
\rho^s(0) = |\alpha^{\prime}, \beta^{\prime} \rangle \langle \alpha^{\prime}, 
\beta^{\prime}|. 
\label{eq:atomcohqnd} 
\end{equation}
Using the Eqs. (\ref{2a.3}), (\ref{eq:atomcohqnd}) in Eq. (\ref{2a.7}) 
we obtain the phase distribution for a two-level
atom, with $j = \frac{1}{2}$ as \cite{sb06} 
\begin{equation}
{\cal P}(\phi) = {1 \over 2 \pi}\left[1 + {\pi \over 4} 
\sin(\alpha^{\prime}) \cos(\beta^{\prime} + \omega t - \phi) e^{- 
(\hbar \omega)^2 \gamma(t)}\right]. \label{2a.9} 
\end{equation}
It can be easily checked that this ${\cal P}(\phi)$ is 
normalized to unity. As can be seen from Eq. (\ref{2a.9}), only 
$\gamma(t)$ plays a role in the effect of the environment on 
the phase distribution.

\subsubsection{System initially in an atomic squeezed state}

Now we consider our system $S$ to be initially in an atomic 
squeezed state \cite{as76, mr78, ds94, ap90} expressed in terms 
of the Wigner-Dicke states as 
\begin{equation}
|\zeta, p \rangle = A_p \exp(\Theta J_z) \exp(-i {\pi 
\over 2} J_y))|j, p \rangle, \label{2a.10}  
\end{equation} 
where
\begin{equation}
e^{2 \Theta} = \tanh(2 |\zeta|) \label{eq:zeta}
\end{equation}
and $A_p$ is usually obtained by normalization.
Thus the initial density matrix of the system $S$ is
\begin{equation}
\rho^s(0)= |\zeta, p \rangle \langle \zeta, p|. \label{2a.12}
\end{equation}
Using the Eqs. (\ref{2a.3}), (\ref{2a.12}) in Eq. (\ref{2a.7}) 
we obtain  the phase distribution for a two-level
atom, with $j = \frac{1}{2}$ for $p=\pm \frac{1}{2}$ 
as \cite{sb06}
\begin{equation}
{\cal P}(\phi) = {1 \over 2 \pi}  
\left[1 \pm \frac{\pi}{4 \cosh(\Theta)} 
\cos(\phi - \omega t ) e^{-(\hbar \omega 
)^2 \gamma(t)} \right],  \label{2a.13} 
\end{equation}
It can be seen that Eqs. (\ref{2a.13}), are normalized to unity.

The  above  expressions  may  be  extended to  the  case  of  multiple
two-level atoms.  For e.g., the  quantum phase
distribution for two two-level atoms, with $j=1$, is: 
\begin{eqnarray}
{\cal P}(\phi) &=& \frac{1}{2\pi}\left\{1 \pm \frac{3\pi}{4(1+\cosh(2\Theta))}
\left[\cos(\phi-\omega t)\cos([\hbar\omega]^2\eta(t))\cosh(\Theta)
\right. \right. \nonumber\\ &-& \left. \left.
\sin(\phi-\omega t)\sin([\hbar\omega]^2\eta(t))\sinh(\Theta)\right]
\exp(-[\hbar\omega]^2\gamma(t)) \right. \nonumber \\ 
&+& \left.  \frac{1}{2(1+\cosh(2\Theta))}\cos(2[\phi-\omega t])
\exp(-4[\hbar\omega]^2\gamma(t))\right\}, \label{2aQND}
\end{eqnarray}
for $p=\pm 1$ and 
\begin{equation}
{\cal P}(\phi) = \frac{1}{2\pi}\left\{1-\frac{1}{2\cosh(2\Theta)}
\cos(2(\phi-\omega t))\exp(-4[\hbar\omega]^2\gamma(t))\right\},
\end{equation}
for $p=0$.

\begin{figure}
\includegraphics{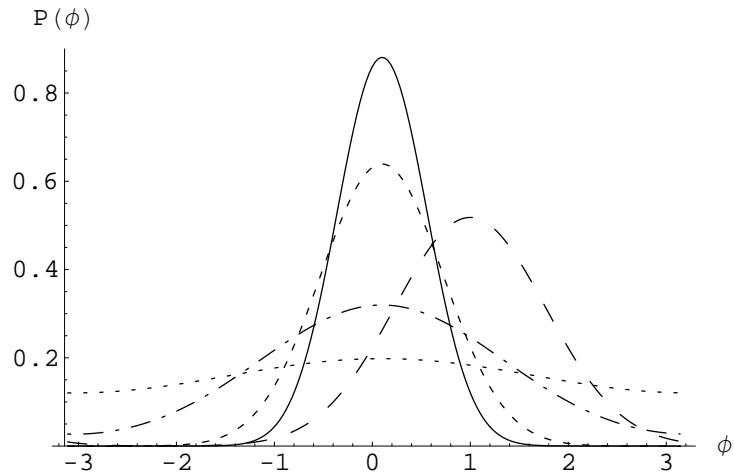}
\caption{Quantum phase distribution ${\cal P}(\phi)$ with respect to
$\phi$ (in radians) for ten atoms, starting in an atomic squeezed
state (Eq. (\ref{2a.12})),
with $j=p=5$, $a=0$ (Eq. (\ref{eq:a})), $\Theta=-0.01832$
(Eq. (\ref{eq:zeta})) and $\gamma_0=0.025$, 
undergoing a QND system-bath interaction. Here $\omega=1$ and $\omega_c=100$.
The bold curve represents unitary evolution for $t=0.1$,
while the small-dashed and large-dashed curves are for the
bath squeezing parameter $r=1.0$, temperature (in units 
where $\hbar\equiv k_B\equiv1$)
$T=0.0$ and evolution times $t=0.1$ and 1, respectively.
The dot-dashed curve represents the case $r=2.0$, $t=0.1$, $T=0.0$,
and the dotted curve the case $r=1.0$, $t=0.1$, $T=300.0$.}
\label{fig:qnd10np}
\end{figure}

In comparison  with Eq. (\ref{2a.13}),  which gives the  quantum phase
distribution  for  a  single  two-level  atom, it  can  be  seen  that
Eq.  (\ref{2aQND})  (phase   distribution  for  two  two-level  atoms)
involves  both  $\eta(t)$  and  $\gamma(t)$.  This  procedure  may  be
carried to any number of two-level atoms using the Wigner-$d$ function
\cite{var88}:
\begin{equation} 
d^j_{n,p}(\pi/2) =2^{-j}\sqrt{(j+n)!(j-n)!(j+p)!(j-p)!)}
\sum_q \frac{(-1)^q}{q!(j+n-q)!(j-p-q)!(p+q-n)!},
\label{eq:wignerd}
\end{equation} 
where $d^j_{n,p}(\theta)$ is the standard Wigner symbol for the
rotation operator \cite{var88}
\begin{equation}
\label{eq:wigsym}
d^j_{n,p}\left(\theta\right) = \langle j,n|e^{-i\theta J_y}|j,p\rangle.
\end{equation}
In Figure  \ref{fig:qnd10np}, we  plot the quantum  phase distribution
for ten two-level atoms.  It can  be clearly seen from the figure that
compared to the unitary case, interaction with the bath (characterized
by finiteness  of $\gamma_0$) causes phase diffusion.  A comparison of
the  small- and large-dashed  curves indicates  that with  increase in
bath exposure duration $t$, the phase distribution diffuses as well as
shifts to the right. It is  also evident from the figure that increase
in  the  bath squeezing  $r$  and  temperature  $T$ also  cause  phase
diffusion. The phase distributions are normalized to unity.

In the case of QND type of interaction, 
the system is decohered without its energy being affected.
This is reflected in the fact that with higher noise,
the `phase' gets completely randomized, resulting
in a flattening of the distribution $P(\phi)$, as depicted 
in Figure \ref{fig:qnd10np}, whereas the `number' distribution,
given by
\begin{eqnarray}
p(m) &=& \langle j,m|\rho^s(t)|j,m\rangle,\hspace*{0.5cm}|m| \le j,\\
 &=& \left\{ \begin{array}{ll}
\left( \begin{array}{c} 2j \\ j+m\end{array}\right) 
(\sin(\alpha^{\prime}/2))^{2(j+m)}
(\cos(\alpha^{\prime}/2))^{2(j-m)} & ~~
{\rm~for~initial~atomic~coherent~state~Eq.
~(\ref{eq:atomcohqnd})} \\
~ & ~\\
|A_p|^2 e^{2m\Theta} |d^j_{mp}(\pi/2)|^2 
& ~~{\rm~for~initial~atomic~squeezed~state~Eq.
~(\ref{2a.12}),} 
\end{array} \right.
\label{eq:pm}
\end{eqnarray}
remains  unaffected. The  distributions  $p(m)$ and  $P(\phi)$ may  be
thought  of as  complementary \cite{as96}  in the  sense  of conjugate
Hermitian observables. For example, it may be verified that a `number'
state, i.e.,  Wigner-Dicke state, corresponds to  a phase distribution
of maximum  uncertainty (in the entropic  sense) \cite{entphas}.  This
process may  be understood as the  selection of states  in a preferred
pointer  basis   \cite{wz93,  mp98},  which  in  this   case  are  the
Wigner-Dicke  states, because  of the  nature of  the system-reservoir
interaction,  whereby the  environment  `monitors' the  system in  the
preferred basis.   As $p(m)$  represents information in  the preferred
basis  \cite{mp98},  the influence  of  the  environment  is not  seen
explicitly in Eq. (\ref{eq:pm}).

\subsection{Harmonic Oscillator System}

Here the system of interest $S$ is taken to be a harmonic oscillator  with 
the Hamiltonian 
\begin{equation} 
H_S = \hbar \omega \left(a^{\dag}a + {1 \over 2} \right).
\label{2b.1}
\end{equation}
The number states serve as an appropriate basis for 
the system Hamiltonian and the system energy eigenvalue 
(\ref{2b.1}) in this basis is 
\begin{equation}
E_{n}= \hbar \omega(n + {1 \over 2}). \label{2b.2} 
\end{equation}
Following Agarwal {\it et al.} \cite{as92} we define a phase 
distribution ${\cal P}(\theta)$ for a given density operator 
$\hat{\rho}$ associated with a state $|\theta \rangle$ as 
\begin{eqnarray}
{\cal P}(\theta) &=& {1 \over 2\pi} \langle \theta|\rho| \theta 
\rangle, ~ 0 \leq \theta \leq 2\pi, \nonumber\\ &=& {1 \over 
2\pi} \sum\limits_{m, n=0}^{\infty} \rho_{m, n} e^{ i(n-
m)\theta},  \label{2b.3} 
\end{eqnarray}
where the states $|\theta\rangle$ are the analogues of the 
Susskind-Glogower \cite{sg64} phase operator and are defined 
in terms of the number states $|n\rangle$ as 
\begin{equation}
|\theta\rangle = \sum\limits_{n=0}^{\infty} e^{i n \theta} 
|n\rangle. \label{2b.4} 
\end{equation}
The sum in Eq. (\ref{2b.3}) is assumed to converge and the phase 
distribution normalized to unity. 
Now we take up two physically interesting initial conditions for the
system $S$.

\subsubsection{System initially in a coherent state}

The initial density matrix of the system is
\begin{equation}
\rho^s(0) = |\alpha \rangle \langle\alpha|, \label{2b.5}
\end{equation}
where 
\begin{equation}
\alpha = |\alpha| e^{i \theta_0} \label{2b.6}
\end{equation}
is a coherent state \cite{sz97}.
Using Eqs. (\ref{2b.2}), (\ref{2b.5}) in Eq. (\ref{2.2}) and then using
it in Eq. (\ref{2b.3}), the phase distribution is obtained as \cite{sb06}
\begin{eqnarray}
{\cal P}(\theta) &=& {1 \over 2\pi} \sum\limits_{m, 
n=0}^{\infty} {|\alpha|^{n+m} \over \sqrt{(n)!(m)!}} e^{i (n-m) 
(\theta - \theta_0)}  e^{-
|\alpha|^2} \nonumber\\ &\times& e^{-i \omega(m-n)t} e^{i (\hbar 
\omega)^2 (m-n)(n+m+1) \eta(t)} e^{-(\hbar \omega)^2 (n-m)^2 
\gamma(t)}. \label{2b.7} 
\end{eqnarray}

\subsubsection{System initially in a squeezed coherent state}

The initial density matrix of the system is
\begin{equation}
\rho^s(0) = |\xi, \alpha \rangle \langle\alpha, \xi|, 
\label{eq:sqcoho} 
\end{equation}
where the squeezed coherent state is defined as \cite{sz97} 
\begin{equation} 
|\xi, \alpha \rangle = S(\xi) D(\alpha) |0\rangle. \label{2b.9}
\end{equation}
Here $S$ denotes the standard squeezing operator and $D$ 
denotes the standard displacement operator \cite{sz97}.
Using Eqs. (\ref{2b.2}), (\ref{eq:sqcoho}) in Eq. (\ref{2.2}) and then using
it in Eq. (\ref{2b.3}), the phase distribution is obtained as \cite{sb06}
\begin{eqnarray}
{\cal P}(\theta) &=& {1 \over 2\pi} \sum\limits_{m, 
n=0}^{\infty} e^{i(n-m)\theta} {e^{i{\psi \over 2}(m-n)} \over 
2^{(m+n) \over 2} \sqrt {(m)!(n!)}} {(\tanh(r_1))^{(m+n) \over 2} 
\over \cosh(r_1)} \nonumber\\ & & \times \exp \left[-|\alpha|^2 
(1 - \tanh(r_1) \cos(2\theta_0 - \psi)) \right] \nonumber\\ & & 
\times H_m \left[{|\alpha| e^{i(\theta_0 - {\psi \over 2})} 
\over \sqrt{\sinh(2r_1)}} \right] H^{*}_n \left[{|\alpha| 
e^{i(\theta_0 - {\psi \over 2})} \over \sqrt{\sinh(2r_1)}} 
\right] \nonumber\\ & & \times  e^{-i \omega(m-n)t} e^{i (\hbar 
\omega)^2 (m-n)(n+m+1) \eta(t)} e^{-(\hbar \omega)^2 (n-m)^2 
\gamma(t)}. \label{2b.10} 
\end{eqnarray} 
Here  the  system squeezing  parameter  $\xi  =  r_1 e^{i  \psi}$  and
$H_n[z]$ is a Hermite  polynomial. The phase distributions depicted by
Eqs.    (\ref{2b.7}),    (\ref{2b.10})    have   been    plotted    in
Ref. \cite{sb06},  where they were  seen to exhibit a  phase diffusion
pattern with the phase distributions being normailzed to unity.

\section{Quantum Phase Distribution of a Two-Level Atomic System in Non-QND
Interaction with Bath}

Here  we will  obtain the  quantum phase  distribution of  a two-level
atomic system in an interaction with a squeezed thermal bath such that
it undergoes  both decoherence  and dissipation.  The  reduced density
matrix operator of the system $S$ is given by \cite{sz97, bp02}
\begin{eqnarray}
{d \over dt}\rho^s(t) &=& -i \frac{\omega}{2}\left[\sigma_z , 
\rho^s (t) \right] \nonumber\\ 
& + & \gamma_0 (N + 1) \left(\sigma_-  \rho^s(t)
\sigma_+ - {1 \over 2}\sigma_+ \sigma_- \rho^s(t) -
{1 \over 2} \rho^s(t) \sigma_+ \sigma_- \right) \nonumber\\
& + & \gamma_0 N \left( \sigma_+  \rho^s(t)
\sigma_- - {1 \over 2}\sigma_- \sigma_+ \rho^s(t) -
{1 \over 2} \rho^s(t) \sigma_- \sigma_+ \right) \nonumber\\
& - & \gamma_0 M   \sigma_+  \rho^s(t) \sigma_+ -
\gamma_0 M^* \sigma_-  \rho^s(t) \sigma_-. \label{3a} 
\end{eqnarray}

In the context of quantum information, the open system effect depicted
by Eq. (\ref{3a})  can be modeled by a  familiar noisy channel called
the  generalized amplitude  damping  channel \cite{bsri06,  deleter,
nc00} for zero bath squeezing.   For the case of finite bath squeezing
and temperature, the corresponding  noisy channel has been obtained by
us  recently  \cite{srb07}  and  could  appropriately  be  called  the
squeezed generalized amplitude damping channel.

In Eq. (\ref{3a}), $\gamma_0$ , having the dimension of $({\rm time})^{-1}$,
is the spontaneous emission rate given by
\begin{equation}
\gamma_0 = {4 \omega^3 |\vec{d}|^2 \over 3 \hbar c^3},
\end{equation}
and $\sigma_+$, $\sigma_-$ are the standard raising and lowering operators,
respectively given by
\begin{eqnarray}
\sigma_+ &=& |1 \rangle \langle 0| =  {1 \over 2}
\left(\sigma_x + i \sigma_y \right), \nonumber\\ 
\sigma_- &=& |0 \rangle \langle 1| = {1 \over 2}
\left(\sigma_x - i \sigma_y \right),  \label{3b}
\end{eqnarray}
with $\sigma_z$ being the standard Pauli operator related to the raising
and lowering operators as $\left[\sigma_+ , \sigma_- \right] = \sigma_z$.
In the above equations, $\left[a, b \right] = ab - ba$.
In Eq. (\ref{3a}) 
\begin{equation}
N = N_{\rm th}(\cosh^2(r) + \sinh^2(r)) + \sinh^2(r), \label{3c} 
\end{equation}
\begin{equation}
M = -\frac{1}{2} \sinh(2r) e^{i\Phi} (2 N_{\rm th} + 1)
\equiv Re^{i\Phi}, \label{3d}
\end{equation}
and
\begin{equation}
N_{\rm th} = {1 \over e^{{\hbar \omega \over k_B T}} - 1}. \label{3e}
\end{equation}
Here $N_{\rm th}$ is the Planck distribution giving the number of thermal
photons at the frequency $\omega$ and $r$, $\Phi$ are squeezing parameters.
The analogous case of a thermal bath without squeezing can be obtained
from the above expressions by setting these squeezing parameters to zero.
Eq. (\ref{3a}) can be solved using the Bloch vector formalism (cf. 
\cite{bp02}, \cite{bsri06}). However, the solutions obtained thus are not
amenable to treatment of the quantum phase distribution by use of 
Eq. (\ref{2a.7}). For this purpose we briefly detail the solution of
Eq. (\ref{3a}) in an operator form. We closely follow the derivation 
given by Nakazato et al. \cite{nh06} and extend it to the case of a 
squeezed thermal bath.

The Eq. (\ref{3a}) can be written as
\begin{equation}
\frac{d}{dt}\rho^s (t) = A \rho^s (t) + \rho^s (t) A^{\dag} +
\left[\gamma_+ \sigma_- \rho^s (t) \sigma_+ + \gamma_- \sigma_+ \rho^s (t) 
\sigma_- - \gamma_0 M \sigma_+ \rho^s (t) \sigma_+
- \gamma_0 M^* \sigma_- \rho^s (t) \sigma_- \right], \label{3f}
\end{equation}
where
\begin{equation}
\gamma_+ = \gamma_0 (N + 1), ~ \gamma_- = \gamma_0 N, \label{3g}
\end{equation}
and
\begin{eqnarray}
A &=& - \frac{1}{4} \gamma^{\beta} - \frac{1}{4} (\gamma + 2 i \omega) 
\sigma_z, \nonumber\\
\gamma^{\beta} &=& \gamma_+ + \gamma_- = \gamma_0 (2 N + 1), \nonumber\\
\gamma &=& \gamma_+ - \gamma_- = \gamma_0. \label{3h}   
\end{eqnarray}
The following transformation is now introduced in Eq. (\ref{3f}):
\begin{equation}
\rho^s (t) = e^{A t} \rho^{I} (t) e^{A^{\dag} t}, \label{3i}
\end{equation}
yielding
\begin{equation}
\frac{d}{dt}\rho^I (t) = \gamma_+ \sigma_- \rho^I (t) \sigma_+ e^{-\gamma t}
+ \gamma_- \sigma_+ \rho^I (t) \sigma_- e^{\gamma t}
- \gamma_0 M \sigma_+ \rho^I (t) \sigma_+ e^{i2\omega t}
- \gamma_0 M^* \sigma_- \rho^I (t) \sigma_- e^{-i2\omega t}.\label{3j}
\end{equation}
The solution of Eq. (\ref{3j}) is facilitated by the introduction of 
superoperators having the following action:
\begin{eqnarray}
{\cal P}_- \rho &=& \sigma_- \rho \sigma_+, ~ 
{\cal P}_+ \rho = \sigma_+ \rho \sigma_-, \nonumber\\
{\cal P}^a_- \rho &=& \sigma_- \rho \sigma_-, ~
{\cal P}^a_+ \rho = \sigma_+ \rho \sigma_+. \label{3k}
\end{eqnarray}
Using Eqs. (\ref{3k}), Eq. (\ref{3j}) can be written as
\begin{eqnarray}
\frac{d}{dt}\rho^I (t) &=& \left[\gamma_+ e^{-\gamma t} {\cal P}_-
+ \gamma_- e^{\gamma t} {\cal P}_+ \right] \rho^I (t) \nonumber\\
&-& \left[\gamma_0 M e^{i2\omega t} {\cal P}^a_+
+ \gamma_0 M^* e^{-i2\omega t} {\cal P}^a_- \right] \rho^I (t). \label{3l}
\end{eqnarray}
Integrating we get
\begin{eqnarray}
\rho^I (t) &=& \rho^I (0) + \frac{1}{\gamma^{\beta}} \left[
\gamma_+ (e^{\gamma_- t} - e^{- \gamma_+ t}) {\cal P}_- +
(\gamma_+ e^{\gamma_- t} + \gamma_- e^{-\gamma_+ t} - \gamma^{\beta})
{\cal P}_- {\cal P}_+ \right] \rho^I (0) \nonumber\\
&+& \frac{1}{\gamma^{\beta}} \left[
\gamma_- (e^{\gamma_+ t} - e^{- \gamma_- t}) {\cal P}_+ +
(\gamma_- e^{\gamma_+ t} + \gamma_+ e^{-\gamma_- t} - \gamma^{\beta})
{\cal P}_+ {\cal P}_- \right] \rho^I (0) \nonumber\\
&-& \gamma_0 M \left[\frac{\sinh(\alpha t)}{\alpha} e^{i \omega t}
{\cal P}^a_+ - \frac{1}{\gamma_0 M} (e^{i \omega t} \{\cosh(\alpha t)
- \frac{i \omega}{\alpha}\sinh(\alpha t)\} - 1 ) {\cal P}^a_+ 
{\cal P}^a_- \right] \rho^I (0) \nonumber\\
&-& \gamma_0 M^* \left[\frac{\sinh(\alpha t)}{\alpha} e^{-i \omega t}
{\cal P}^a_- - \frac{1}{\gamma_0 M^*} (e^{-i \omega t} \{\cosh(\alpha t)
+ \frac{i \omega}{\alpha}\sinh(\alpha t)\} - 1 ) {\cal P}^a_- 
{\cal P}^a_+ \right] \rho^I (0), \label{3m}
\end{eqnarray}
where 
\begin{equation}
\alpha = \sqrt{\gamma^2_0 |M|^2 - \omega^2}. \label{3n}
\end{equation}
All the other terms are as given above.
Using Eq. (\ref{3i}) in Eq. (\ref{3m}) we finally obtain the solution of
Eq. (\ref{3a}) as
\begin{eqnarray}  
\rho^s (t) &=& \frac{1}{4} \rho^s (0) (1 + e^{- \gamma^{\beta} t} + 
2 \cosh(\alpha t) e^{-\frac{\gamma^{\beta} t}{2}}) \nonumber\\
&+& \frac{1}{4} \sigma_z \rho^s (0) \sigma_z (1 + e^{- \gamma^{\beta} t} - 
2 \cosh(\alpha t) e^{-\frac{\gamma^{\beta} t}{2}}) \nonumber\\
&-& \frac{1}{4} \rho^s (0) \sigma_z \left( \frac{\gamma}{\gamma^{\beta}}
(1 - e^{- \gamma^{\beta} t}) - \frac{2 i \omega}{\alpha} \sinh(\alpha t)
e^{-\frac{\gamma^{\beta} t}{2}} \right) \nonumber\\
&-& \frac{1}{4} \sigma_z \rho^s (0)  \left( \frac{\gamma}{\gamma^{\beta}}
(1 - e^{- \gamma^{\beta} t}) + \frac{2 i \omega}{\alpha} \sinh(\alpha t)
e^{-\frac{\gamma^{\beta} t}{2}} \right) \nonumber\\
&+& (1 - e^{- \gamma^{\beta} t}) \left(\frac{\gamma_+}{\gamma^{\beta}}
 \sigma_- \rho^s (0) \sigma_+ + \frac{\gamma_-}{\gamma^{\beta}}
 \sigma_+ \rho^s (0) \sigma_- \right) \nonumber\\
&-& \gamma_0 \frac{\sinh(\alpha t)}{\alpha} e^{-\frac{\gamma^{\beta} t}{2}}
\left(M \sigma_+ \rho^s (0) \sigma_+ 
+ M^* \sigma_- \rho^s (0) \sigma_- \right). \label{3o}
\end{eqnarray}
This is the desired form of solution of the master equation (\ref{3a}). For 
the case of a thermal bath without squeezing, $r$ and $\Phi$ are zero and
it can be seen that Eq. (\ref{3o}) reduces to the solution obtained by
Nakazato et al. \cite{nh06} for the case of a two-level atom interacting
with a thermal bath. We will use Eq. (\ref{3o}) in the following
subsections
to investigate the quantum phase distribution.  

\subsection{System initially in an atomic coherent state}

Taking the  intial density matrix of the system $S$  to be as in
Eq. (\ref{eq:atomcohqnd}), using it in Eq. (\ref{3o}),  and then in
Eq. (\ref{2a.7}), with $j =  \frac{1}{2}$, we obtain the quantum phase
distribution as
\begin{eqnarray}
{\cal  P}(\phi) &=&  \frac{1}{2  \pi} \left[1  + \frac{\pi}{4  \alpha}
\sin(\alpha^{\prime})   \Big\{\alpha    \cosh(\alpha   t)\cos(\phi   -
\beta^{\prime}) +  \omega \sinh(\alpha t)  \sin(\phi - \beta^{\prime})
\right. \nonumber\\ & & \left.  - \gamma_0 R \sinh(\alpha t) \cos(\Phi
+  \beta^{\prime}  +   \phi)  \Big\}  e^{-\frac{\gamma^{\beta}  t}{2}}
\right].  \label{3p}
\end{eqnarray}
Here $R$, $\Phi$ come from Eq. (\ref{3d}),
and $\gamma^{\beta}$, $\alpha$ are as in Eqs. (\ref{3h}) and (\ref{3n}),
respectively. The Eq. (\ref{3p}) can be seen to be normalized to unity.
When $\gamma_0$ is set equal to zero, i.e., for the case where 
the effects of the bath are neglected, Eq. (\ref{3p}) becomes
\begin{equation}
{\cal P}(\phi, \gamma_0 = 0) = \frac{1}{2\pi}\{1 + \frac{\pi}{4}
\sin(\alpha^{\prime}) \cos(\beta^{\prime} + \omega t - \phi) \}. \label{3q}
\end{equation}
In the analogous case of the QND system-bath interaction,
the phase distribution was given by Eq. (\ref{2a.9}) which with
the bath coupling
parameter $\gamma_0$ set to zero,  is easily seen to reduce to
Eq. (\ref{3q}). This is a nice consistency check for these equations.

\begin{figure}
\includegraphics{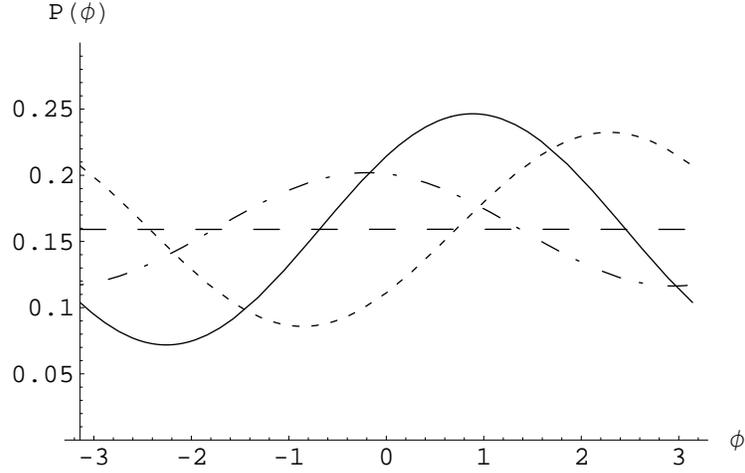}
\caption{Quantum phase distribution ${\cal P}(\phi)$ (Eq. (\ref{3p}))
with respect to $\phi$ (in radians), for 
a two-level dissipative system initially in an atomic coherent state
(\ref{eq:atomcohqnd}). 
Here $\omega=1.0$, $\Phi=\pi/8$, $\alpha^{\prime}
=\beta^{\prime}=\pi/4$, and  $\gamma_0 = 0.25$.
The bold and small-dashed curves correspond to 
temperature (in units where $\hbar\equiv k_B\equiv1$) $T=0$, 
bath squeezing parameter $r=0$, but with bath exposure times $t=0.1$
and $1.5$, respectively.
The large-dashed and dot-dashed curves 
correspond to $T=300$ and $t=0.1$, but $r=0$ and $2$, respectively.
Comparing the last two curves, we note that, counterintuitively,
squeezing resists diffusion.}
\label{fig:atomCohDis}
\end{figure}

Figure  \ref{fig:atomCohDis}  illustrates   the  combined  effects  of
temperature, evolution  time and bath squeezing  ($r,\Phi$) on quantum
phase distribution.  Comparison of  the small- and large-dashed curves
brings out the diffusive  influence of temperature, while a comparison
of the bold and small-dashed  curves shows that the phase distribution
shifts  with increase  in bath  exposure time.  On the  other  hand, a
comparison  between the  large- and  dot-dashed curves  illustrates an
interesting feature  of squeezing  in dissipative systems  governed by
Lindblad-type  equations   (\ref{3a}),  in  that   squeezing  tends  to
counteract  the  influence of  temperature,  which  in  this case 
manifests as resistence to  randomization of phase.   A similar
behavior is observed in the  joint effect of temperature and squeezing
on  the geometric  phase  of a  qubit  (two-level system)  interacting
dissipatively with its environment \cite{bsri06}. The normalization of
the phase distribution is preserved.

\begin{figure}
\includegraphics{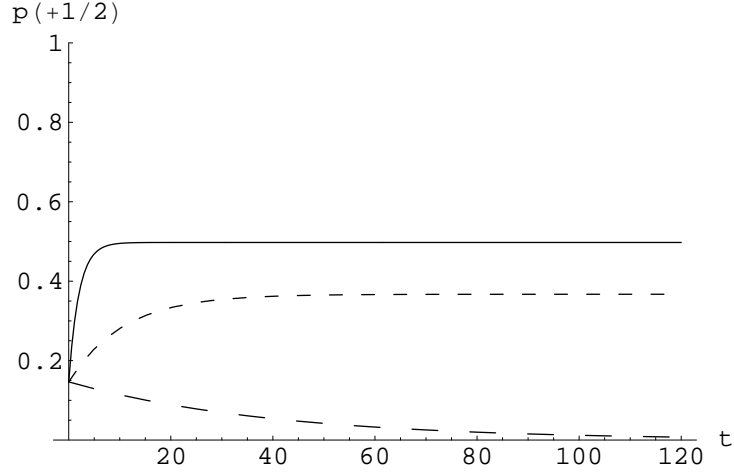}
\caption{The  distribution  $p(m=1/2,t)$  (Eq. (\ref{eq:pmc}))  for  a
two-level  dissipative system  starting  in an  atomic coherent  state
(Eq.  (\ref{eq:atomcohqnd}),  as  a  function of  time  for  different
environmental  conditions.   The  bold  curve corresponds  to 
temperature $T=100$, 
$\gamma_0=0.0025$, $r=\Phi=0$, $\omega=1$, $\alpha^{\prime}=
\beta^{\prime}=\pi/4$, illustrative  of  a system  becoming  maximally
mixed  with  time.   The   large-dashed  curve  corresponds  to  $T=0,
\gamma_0=0.025,r=0$, and depicts quantum deletion \cite{deleter}.  The
small-dashed curve represents the case $T=0, \gamma_0=0.025,r=1$.
Here time and temperature are in units where $\hbar\equiv k_B\equiv1$.}  
\label{fig:atomcohpm}
\end{figure}

We plot in Figure \ref{fig:atomcohpm} the function
\begin{eqnarray}
p(m=1/2,t) &=& \langle 1/2|\rho^s(t)|1/2\rangle \nonumber \\
&=& \frac{1}{2}\left[\left(1-\frac{\gamma_0}{\gamma^\beta}\right)
+ \left(1+\frac{\gamma_0}{\gamma^\beta}\right)e^{-\gamma^\beta t}\right]
\sin^2(\alpha^\prime/2) +
\frac{\gamma_-}{\gamma^\beta}\left(1-e^{-\gamma^\beta t}\right)
\cos^2(\alpha^\prime/2). 
\label{eq:pmc}
\end{eqnarray}

\begin{figure}
\includegraphics{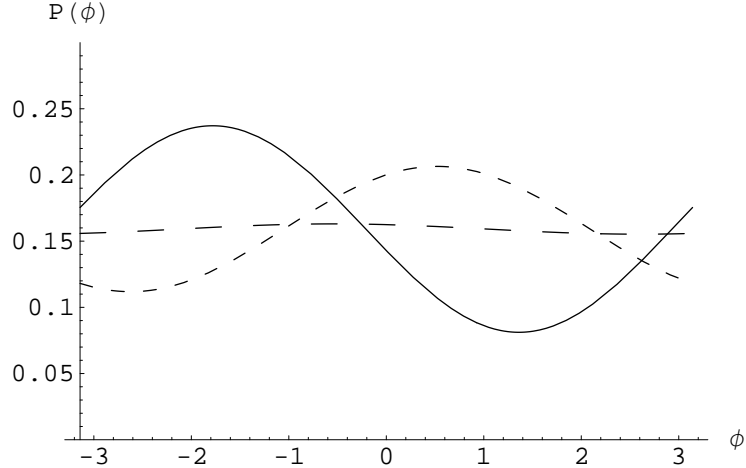}
\caption{Quantum phase distribution  ${\cal P}(\phi)$ (Eq. (\ref{3p}))
with  respect to  $\phi$  (in radians),  for  a two-level  dissipative
system      starting     in      an     atomic      coherent     state
(Eq. (\ref{eq:atomcohqnd}), at various  times with temperature (in
units   where  $\hbar\equiv   k_B\equiv1$)  $T=0$ and
bath squeezing parameters $r=\Phi=0$,
$\gamma_0=0.025$, $\omega  = 1, \alpha^{\prime}=\beta^{\prime}=\pi/4$.
The   large-dashed,   small-dashed   and   bold   curves   correspond,
respectively,  to  evolution  times   $t=250$,  $50$  and  $10$.   The
large-dashed curve depicts the  randomization of phase distribution at
long times.  Comparison of  this figure with Figure  \ref{fig:atomcohpm}
clearly brings  out complementarity  between the `number'  and `phase'
variables. In particular,  comparision between the large-dashed curves
in  both Figures  shows  how  as the  state  becomes increasing  pure,
tending  to  $m=-1/2$,  with  time,  the  corresponding  complementary
distribution $P(\phi)$ level outs.}
\label{fig:atomcohPfi}
\end{figure}

Figure \ref{fig:atomcohpm} depicts an expected behavior of a two-level
system subjected  to a dissipative channel.  In  particular, for $T=0$
and $r=0$, it  drives the system towards a  pure state (with $m=-1/2$)
and thus behaves as a quantum deleter \cite{deleter}. Correspondingly,
the  phase  distribution $P(\phi)$  tends to level out
for large bath exposure time  $t$, as seen in
Figure  \ref{fig:atomcohPfi}.   This
brings  out  nicely  the   complementarity between  $p(m)$  and  $P(\phi)$
\cite{as96}. It is to be noted that, in contrast to the QND case, here
the Wigner-Dicke  states are  not the preferred  basis, and  hence the
environmental effects  manifest themselves  in the function  $p(m)$ as
seen in Eq. (\ref{eq:pmc}).

\subsection{System initially in an atomic squeezed state}

Taking the intial density matrix of the system $S$ to be as in
Eq. (\ref{2a.12}), using it in Eq. (\ref{3o}), and then in Eq. (\ref{2a.7}),
with $j = \frac{1}{2}$, we obtain the quantum phase distribution 
for $p = \pm \frac{1}{2}$ as
\begin{eqnarray}
{\cal P}(\phi) &=& \frac{1}{2\pi}
\left[1 \pm \frac{\pi}{4 \cosh(\Theta)} 
\Big\{\cosh(\alpha t) \cos(\phi) + \frac{\omega}{\alpha}
\sinh(\alpha t) \sin(\phi) % \right. \nonumber\\
% & & \left. 
-\frac{\gamma_0 R}{\alpha} \sinh(\alpha t)
\cos(\phi + \Phi)\Big\} e^{-\frac{\gamma^{\beta} t}{2}} \right].  \label{3r}
\label{eq:spnp}
\end{eqnarray}
Here $\Theta$ is as defined in Eq. (\ref{eq:zeta}) and
all the other terms are as given above. The Eqs. (\ref{eq:spnp}) are 
easily seen to be normalized to unity. Also by setting $\gamma_0$ to zero in 
them, they are seen to reduce to the cases of $\gamma_0$ set to zero in their 
QND counterparts, Eqs. (\ref{2a.13}), respectively. This serves
as a consistency check for these equations.
On comparing the above equations, for the quantum phase distributions, with the
corresponding ones for the case of QND system-bath interaction, these are 
easily seen to be more complicated. This is a reflection of the fact that
the phase distributions developed in this section are for a process 
that involves both dephasing as well as dissipation, 
in contrast to the QND case, which involves only dephasing.
 
\begin{figure}
\subfigure[]{\includegraphics[width=7.0cm]{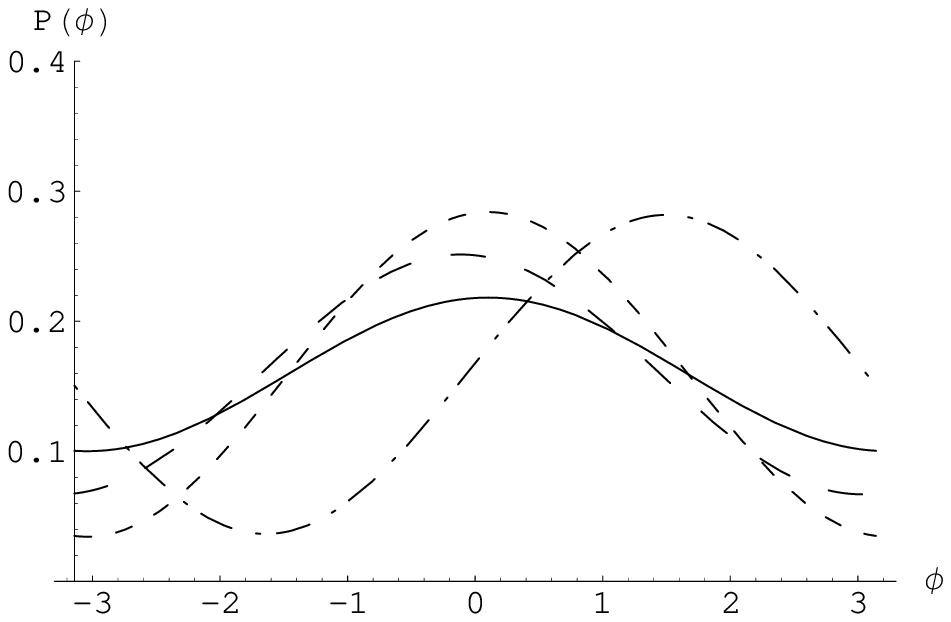}} \hfill
\subfigure[]{\includegraphics[width=7.0cm]{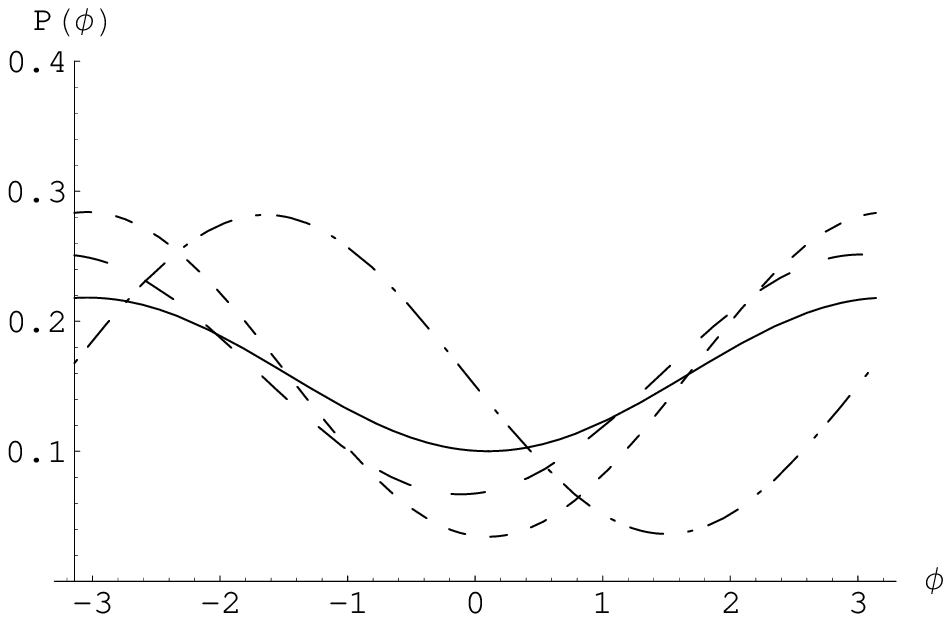}}
\caption{Quantum phase distribution ${\cal P}(\phi)$ (Eq. (\ref{eq:spnp})
with respect to $\phi$ (in radians) for a two-level system starting
in an atomic squeezed state (\ref{2a.12}).
Here $\omega=1.0,\Phi=\pi/8,\Theta=-0.01832$ and
$\gamma_0 = 0.025$. Figure (a) refers to $p=\frac{1}{2}$
and (b) to  $p=-\frac{1}{2}$. 
In both figures, the large-dashed and bold curves 
correspond to temperature 
(in units where $\hbar\equiv k_B\equiv1$) $T=300$ and evolution time $t=0.1$.
The bath squeezing parameter $r$ is, respectively, $0.5$ and $0.0$.
The small-dashed and dot-dashed curves
correspond to $T=0$ and $r=0.0$, with
time $t$ being $0.1$ and $1.5$, respectively.
}
\label{fig:spnp}
\end{figure}

We plot in Figures \ref{fig:spnp} the
quantum  phase  distributions  ${\cal
P}(\theta)$ for a two-level system starting in an atomic squeezed
state (\ref{2a.12}). 
An interesting  feature in  Figures \ref{fig:spnp} is  brought out  by a
comparison of the  bold and large-dashed curves. Squeezing  is seen to
have the  effect of resisting  the diffusive effect of  temperature on
the  phase.   This  is  similar   to  the  behavior  seen   in  Figure
\ref{fig:atomCohDis}, and suggests that  this is a generic property of
squeezing  in   a  dissipative  interaction.   A   comparison  of  the
small-dashed  and  bold curves  brings  out  the  diffusive effect  of
temperature on  the phase distribution while a  comparison between the
small-dashed and dot-dashed curves  shows that the distribution shifts
with time. The phase distribution normalization is preserved.

\section{Quantum Phase Distribution of a Harmonic-Oscillator System in Non-QND
Interaction with Bath}

Here we will obtain the quantum phase distribution of a harmonic-oscillator
system, $H_s = \hbar \omega (a^{\dag} a + \frac{1}{2})$, in a 
dissipative interaction with a squeezed thermal bath.
The reduced density matrix operator of the system $S$,
in the interaction picture, is given by \cite{sz97, bp02}
\begin{eqnarray}
{d \over dt}\rho^s(t) &=& \gamma_0  (N + 1) \left(a \rho^s(t) a^{\dag}
- {1 \over 2} a^{\dag} a \rho^s(t)  - {1 \over 2} \rho^s(t) a^{\dag} a
\right) \nonumber\\ & + & \gamma_0  N \left( a^{\dag} \rho^s(t) a - {1
\over  2}a a^{\dag}  \rho^s(t)  -  {1 \over  2}  \rho^s(t) a  a^{\dag}
\right) \nonumber\\ & + & \gamma_0 M \left(a^{\dag} \rho^s(t) a^{\dag}
- {1  \over  2}(a^{\dag})^2  \rho^s  (t)  - {1  \over  2}  \rho^s  (t)
(a^{\dag})^2 \right) \nonumber\\ &  + & \gamma_0 M^* \left(a \rho^s(t)
a  - {1  \over 2}  (a)^2 \rho^s  (t) -  {1 \over  2} \rho^s  (t) (a)^2
\right). \label{4a}
\end{eqnarray}
In the above equation, $N$, $M$ are bath parameters which will be given
below and $\gamma_0$ is a parameter which depends upon the system-bath coupling
strength. The Eq. (\ref{4a}) can be solved using a variety of methods (cf. 
\cite{bp02}, \cite{sz97}). However, the solutions obtained thus are not
amenable to treatment of the quantum phase distribution by use of 
Eq. (\ref{2b.3}). For this purpose we again briefly detail the solution of
Eq. (\ref{4a}) in an operator form. We closely follow the derivation 
given by Lu et al. \cite{ly03}.
The following transformations are introduced \cite{ek90}:
\begin{equation}
\rho^{'s}(t)  =  S^{\dag}  (\zeta)  \rho^s  (t) S  (\zeta),~  a^{'}  =
S^{\dag} (\zeta) a S (\zeta), \label{4b}
\end{equation} 
where
\begin{equation}
S (\zeta) = e^{\frac{1}{2}(\zeta^* a^2 - \zeta a^{\dag 2})}. \label{4c}
\end{equation}
Using Eqs. (\ref{4b}) we get
\begin{equation}
a^{'}  =  \cosh(|\zeta|)   a  -  \frac{\zeta}{|\zeta|}  \sinh(|\zeta|)
a^{\dag}. \label{4d}
\end{equation}
Using Eqs. (\ref{4b}), (\ref{4d}) in Eq. (\ref{4a}), we get
\begin{equation}
{d \over dt}\rho^{'s}(t) = \left[\alpha K_+ + \beta K_- + (\alpha +
\beta) K_0 + \frac{\gamma_0}{2} \right] \rho^{'s}(t), \label{4e} 
\end{equation}
where
\begin{eqnarray}
\alpha = \gamma_0 N \cosh(2|\zeta|) + \gamma_0 \cosh^2(|\zeta|)
- \frac{\gamma_0}{2|\zeta|} \sinh(2|\zeta|) (M \zeta^* 
+ M^* \zeta), \nonumber\\
\beta = \gamma_0 N \cosh(2|\zeta|) + \gamma_0 \sinh^2(|\zeta|)
- \frac{\gamma_0}{2|\zeta|} \sinh(2|\zeta|) (M \zeta^* 
+ M^* \zeta). \label{4f} 
\end{eqnarray}
The parameters involved in the above equation need to satisfy the following
consistency condition:
\begin{equation}
\frac{|\zeta|}{\zeta} M \coth(|\zeta|) + \frac{\zeta}{|\zeta|} 
M^* \tanh(|\zeta|) = 2 N + 1. \label{4g}
\end{equation}
It can be seen that
\begin{eqnarray}
M &=& \frac{1}{2} \sinh(2r) (2 N_{\rm th} + 1)e^{i\Phi}, \nonumber\\
N &=& N_{\rm th} (\cosh^2 (r) + \sinh^2 (r)) + \sinh^2 (r), \nonumber\\
N_{\rm th} &=& \frac{1}{e^{\frac{\hbar \omega}{k_B T}} - 1},~
\zeta = r e^{i \Phi}, \label{4h} 
\end{eqnarray}
satisfy Eq. (\ref{4g}). In Eq. (\ref{4e}), $K_+$, $K_-$ and $K_0$ are
superoperators satisfying 
\begin{equation}
K_+ \rho^{'s} = a \rho^{'s} a^{\dag},~ K_- \rho^{'s} = a^{\dag} \rho^{'s} a,
K_0 \rho^{'s} = -\frac{1}{2} (a^{\dag} a \rho^{'s} + \rho^{'s} a^{\dag} 
a + \rho^{'s}). \label{4i}
\end{equation}
These superoperators can be seen to satisfy:
\begin{equation} 
\left[K_-, K_+ \right]\rho^{'s}  = 2 K_0 \rho^{'s},~
\left[K_0, K_{\pm} \right]\rho^{'s}  = {\pm} K_{\pm} \rho^{'s}, \label{4j}
\end{equation}
which coincides  with the commutation  relations of the  $su(1,1)$ Lie
algebra.   This  brings  out   the  intimate  connection  between  the
solutions of the master equation  (\ref{4a}) and the generators of the
$su(1,1)$  Lie  algebra.  Using  the  disentangling  theorems  of  the
$su(1,1)$ Lie algebra, Eq. (\ref{4e}) can be solved to yield:
\begin{equation}
\rho^{'s} (t) = e^{\frac{\gamma_0 t}{2}} e^{y_- (t) K_-} 
e^{\ln(y_0 (t)) K_0} e^{y_+ (t) K_+} \rho^{'s} (0), \label{4k}
\end{equation}
where
\begin{eqnarray}
y_0 (t) &=& \left(\frac{\alpha e^{\frac{\gamma_0 t}{2}} - \beta
e^{-\frac{\gamma_0 t}{2}}}{\gamma_0}\right)^2, \nonumber\\
y_+ (t) &=& \frac{\alpha (e^{- \gamma_0 t} -1)}{(\beta e^{- \gamma_0 t}
- \alpha)}, \nonumber\\
y_- (t) &=& \frac{\beta (e^{- \gamma_0 t} -1)}{(\beta e^{- \gamma_0 t}
- \alpha)}. \label{4l} 
\end{eqnarray}
Using Eqs. (\ref{4k}), (\ref{4b}), the solution of Eq. (\ref{4a}) can
be written as 
\begin{equation} 
\rho^s (t) = S(\zeta) \Big\{ e^{\frac{\gamma_0 t}{2}} e^{y_- (t) K_-} 
e^{\ln(y_0 (t)) K_0} e^{y_+ (t) K_+} S^{\dag} (\zeta) \rho^s (0)
S (\zeta)\Big\} S^{\dag} (\zeta). \label{4m} 
\end{equation}
This is the form of solution of the master equation which we will use for
investigation of the quantum phase distribution. We will use a special
initial state of the system, the squeezed coherent state,
\begin{equation}
\rho^s (0) = |\zeta, \eta \rangle \langle \eta, \zeta|, \label{4n}
\end{equation}
where
\begin{equation}
|\zeta, \eta \rangle = S(\zeta) D(\eta) |0 \rangle. \label{4o}
\end{equation}
Here $|0  \rangle$ is the vacuum  state and $D(\eta)$  is the standard
displacement operator. Substituting  Eq. (\ref{4n}) in Eq. (\ref{4m}),
the solution  of the  Eq. (\ref{4a}) starting  from the  initial state
(\ref{4n}), following Lu et al. \cite{ly03}, is obtained as
\begin{eqnarray}
\rho^s (t) &=& \frac{1}{(1 + 
\tilde{\beta}(t))} e^{-\tilde{\beta}(t) |\tilde{\eta}(t)|^2}
\sum\limits_{k=0}^{\infty} \left(\frac{\tilde{\beta}(t)}{(1 + 
\tilde{\beta}(t))}\right)^k \frac{1}{k!} \times \nonumber \\
&& \sum_{l,p=0}^k 
\left( \begin{array}{l} k \\ l \end{array}\right)
\left( \begin{array}{l} k \\ p \end{array}\right)
\sqrt{l! p!} (\tilde{\eta}^*(t))^{k-l} 
(\tilde{\eta}(t))^{k-p} 
|\zeta, \tilde{\eta} (t), l \rangle \langle p, 
\tilde{\eta} (t), \zeta|, \label{4p}
\end{eqnarray}
where
\begin{equation}
|\zeta, \tilde{\eta} (t), l \rangle = S(\zeta) |\tilde{\eta} (t), l \rangle = 
S(\zeta) D(\tilde{\eta} (t)) |l \rangle 
, \label{4q}
\end{equation}
and
\begin{equation}
\tilde{\beta} (t) = \frac{\beta}{\gamma_0} (1 - e^{-\gamma_0 t}),~
\tilde{\eta} (t) = \eta \frac{e^{-\frac{\gamma_0 t}{2}}}
{(1 + \tilde{\beta} (t))}, \label{4r}
\end{equation}
where  $\beta$  is given  by  Eq.   (\ref{4f}).   In Eq.   (\ref{4q}),
$D(\tilde{\eta} (t))  = e^{\tilde{\eta} (t)  a^{\dag} - \tilde{\eta}^*
(t)  a}$  and  $D(\tilde{\eta}  (t))  |l  \rangle$  is  known  as  the
generalized coherent state (GCS)  \cite{rs82, sm91} and thus the state
$|\zeta,  \tilde{\eta}  (t), l  \rangle  $  would  be the  generalized
squeezed coherent state (GSCS)  \cite{sm91}. The GCS's were introduced
by  Roy  and  Singh  \cite{rs82},  where they  demonstrated  that  the
harmonic oscillator  possesses an infinite string  of coherent states.
We see from  Eqs.  (\ref{4p}) and (\ref{4n}) that  under the action of
the  master  equation (\ref{4a}),  which  is  of  a Lindblad  kind,  a
harmonic oscillator  starting in a  squeezed coherent state ends  in a
mixture that can be expressed as a sum over GSCS.  Thus the above case
can be thought of as  a concrete physical realization of GSCS.  

This is an example of ultracoherence pertaining to master equations
governing the Lindblad type of evolution such as Eq. (\ref{4a}).
Ultracoherence refers to the structure induced into the Fock space
${\cal F}({\cal H})$, over a finite or infinite dimensional Hilbert
space ${\cal H}$, by the action of all canonical transformations,
both homogeneous (e.g., squeezing operation)
and inhomogeneous (Weyl operators) \cite{bk05,kb06}.
Starting from the squeezed coherent state (\ref{4n})
of the harmonic oscillator, obtained by applying    
the canonical transformation $U=S(\zeta)D(\eta)$ (\ref{4o})
to the vacuum state, and applying
a canonical transformation (\ref{4b}) to the master equation (\ref{4a}),
results in a mixture of ultracoherent
states, which in this case is the GSCS.

Making
use of the Fock-space representation of GCS \cite{rs82}
\begin{equation} 
|n, \alpha(t) \rangle = e^{-\frac{|\alpha(t)|^2}{2}}
\sum\limits_{l=0}^{\infty} \left(\frac{n!}{l!} \right)^{\frac{1}{2}}
L^{l - n}_n (|\alpha(t)|^2) [\alpha(t)]^{l-n} |l \rangle
, \label{4s}
\end{equation}
where $L^{l - n}_n (x)$ is the generalized Laguerre polynomial, and 
substituting Eq. (\ref{4p}) in Eq. (\ref{2b.3}), reverting back to the
Schr$\ddot{o}$dinger picture, we obtain the quantum
phase distribution of a dissipative harmonic oscillator starting in a
squeezed coherent state (\ref{4n}) as
\begin{eqnarray} 
{\cal             P}(\theta)             &=&            \frac{1}{2\pi}
e^{-|\tilde{\eta}(t)|^2}\frac{e^{-\tilde{\beta}(t)|\tilde{\eta}(t)|^2}}
{(1  +  \tilde{\beta}(t))}  \sum\limits_{m,n}  e^{- i  \omega  (m-n)t}
e^{i(n-m)\theta} \sum\limits_{u,v,k} G^*_{u,m} (\zeta) G_{v,n} (\zeta)
\nonumber\\      &\times&      \left(\frac{\tilde{\beta}(t)}{(1      +
\tilde{\beta}(t))}\right)^k   \frac{1}{k!}   \sum_{l,p=0}^{k}   \left(
\begin{array}{l} k  \\ l \end{array}\right)  \left( \begin{array}{l} k
\\     p    \end{array}\right)     \frac{l!     p!}{\sqrt{(u!    v!)}}
(\tilde{\eta}^*(t))^{v-p+k-l}  (\tilde{\eta}(t))^{u-l+k-p}  \nonumber  \\
&\times&   L^{u   -   l}_l   (|\tilde{\eta}(t)|^2)  L^{*{v   -   p}}_p
(|\tilde{\eta}(t)|^2). \nonumber\\
\label{4t}
\end{eqnarray}
In the above equation, $G_{m,n} (\zeta) = \langle m | S(\zeta) | n \rangle$ 
and is explicitly given, with $\zeta = r_1 e^{i \phi}$, as \cite{sm91}
\begin{eqnarray}
G_{2m,2p} &=& {(-1)^{p } \over (p)! (m)!} \left({(2p)! (2m)! 
\over \cosh(r_1)}\right)^{1 \over 2} \exp{\left(i(m - p)\phi 
\right)} \nonumber\\ & & \times \left({\tanh(r_1) \over 2} 
\right)^{(m+p)} F^2_1 \left[-p, -m; {1 \over 2}; -{1 \over 
(\sinh(r_1))^2}\right]. \label{4u} 
\end{eqnarray}
Similarly $G_{2m+1,2p+1}(\zeta)$ is given by
\begin{eqnarray}
G_{2m+1,2p+1} &=& {(-1)^{p } \over (p)! (m)!} \left({(2p+1)! 
(2m+1)! \over \cosh^3(r_1)}\right)^{1 \over 2} \exp{\left(i(m - 
p)\phi \right)} \nonumber\\ & & \times \left({\tanh(r_1) \over 2} 
\right)^{(m+p)} F^2_1 \left[-p, -m; {3 \over 2}; -{1 \over 
(\sinh(r_1))^2}\right]. \label{4v} 
\end{eqnarray}
As has been 
pointed out in \cite{sm91}, $G_{m,n}$ is nonzero only for either $m, 
n$ both even or both odd. For convenience it is sometimes 
assumed that $\phi$ is zero and $z=r_1$ is real. Here $r_1 = r$, due to the
initial condition (\ref{4n}) and $F^2_1$ is 
the Gauss hypergeometric function \cite{ETBM}.

\begin{figure}
\includegraphics{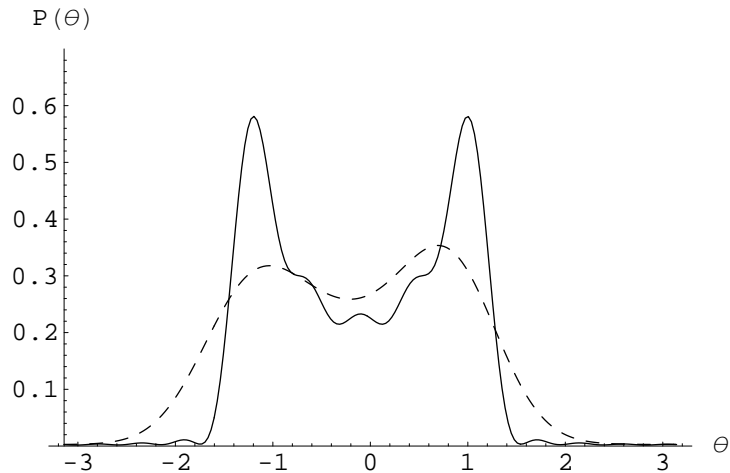}
\caption{A  comparison  of  the  quantum  phase  distributions  ${\cal
P}(\theta)$, for  a harmonic oscillator system starting  in a squeezed
coherent  state, for QND  system-bath interaction  (Eq. (\ref{2b.10}))
with that  for dissipative system-bath  interaction (Eq.  (\ref{4t})).
The former  (latter) is represented  by the dashed (solid)  curve.  In
both cases, temperature  (in units where $\hbar \equiv  k_B \equiv 1$)
$T=0$, the squeezing parameters $r=r_1=1$, bath exposure time $t=0.1$,
$\gamma_0  = 0.025$,  $\omega=1$ .  In the  former case,  $\psi=0$ and
$\omega_c=100$, while in the latter, $\Phi=0$.}
\label{fig:2plots}
\end{figure}
In Figure \ref{fig:2plots}, we make  a comparison of the quantum phase
distributions  ${\cal  P}(\theta)$ for a harmonic
oscillator system starting in a squeezed
coherent state (\ref{2b.9}), for  QND  system-bath  interaction
(Eq. (\ref{2b.10})) with  that for dissipative system-bath interaction
(Eq.   (\ref{4t})).    A  comparison   of  the  distributions 
brings out  the differing  effects  of the  two types  of
system-bath  interactions   on  them.  The  phase
distributions are normalized.

\section{Applications: Phase dispersion \label{sec:appli}}

From  the  perspective of  experiments,  a  relevant  quantity is  the
quantum  phase fluctuation, which  may be  quantified by  the variance
$\sigma^2  =  \langle   \phi^2\rangle  -  \langle\phi\rangle^2$.   For
example, Ref.  \cite{kbm98} presents  measurement of phase variance on
atomic populations  using interferometry improved  by QND measurements
at the inputs  to the interferometer.  However, this  measure of phase
fluctuation  has the drawback  that it  depends on  the origin  of the
phase  integration. A measure  of phase  fluctuation that  avoids this
problem is the dispersion $D$ \cite{pp98,bp69,tom94,lp96},
\begin{equation}
D = 1 - \left|\int_{-\pi}^{+\pi} d\phi e^{-i\phi}{\cal P}(\phi)\right|^2.
\label{eq:var}
\end{equation}
In this section, as an application of the phase distribution formalism
employed  above,  we  study   the  phase  dispersion  $D$  from  these
distributions. We also evaluated  the variance for these distributions
(not presented in  this work), and found that  in certain cases, there
is  in fact  a qualitative  difference of  the behavior  of  these two
quantities.

\begin{figure}
\includegraphics{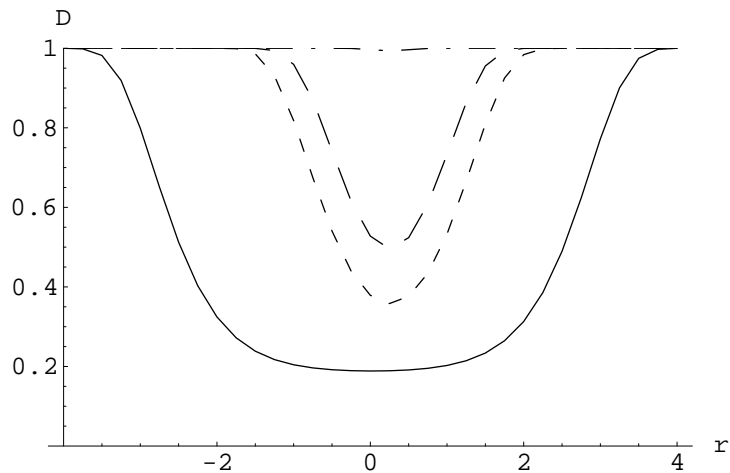}
\caption{Dispersion as  function of environmental  squeezing parameter
$r$, for ten  two-level atomic systems starting in  an atomic squeezed
state  (Eq.   (\ref{2a.12})),  at  various  temperatures   for  a  QND
system-environment  interaction.   Here  $a=0.0$  [Eq.  (\ref{eq:a})],
$\gamma_0=0.0025$,  $\Theta=-0.01832$, $t=1.0$,  $j=p=5$, $\omega=1.0$
and   $\omega_c=100.0$.  The   bold,   small-dashed,  large-dashed   and
dot-dashed  curves correspond  to temperatures  $T$ (in  units where
$\hbar\equiv k_B=1$) 0, 50, 100 and 1000, respectively.}
\label{fig:var1}
\end{figure}

Figure \ref{fig:var1} depicts the behavior of dispersion $D$ of $\phi$
of a  ten two-level  atomic system, starting  from an  atomic squeezed
state, interacting with a squeezed thermal bath via a QND interaction,
with    respect    to    environmental   squeezing    parameter    $r$
(Eq.  (\ref{eq:a})).   The  dispersion   is  found  to  increase  with
temperature  and  squeezing, tending  to  the  maximal  value of  $1$,
corresponding  to  the  uniform  distribution  $P(\phi)=1/2\pi$.  This
indicates that for a QND type system-bath interaction both temperature
and squeezing have a similar effect of causing diffusion of the phase.
Increasing bath exposure time $t$ also leads to the effect of leveling
out $D$ to  $1$. At $T=0$, this leveling out takes  a much longer time
on   account   of    the   logarithmic   dependence   of   $\gamma(t)$
(Eq. (\ref{2.7})) on $t$, indicating a power-law decay.

\begin{figure}
\includegraphics{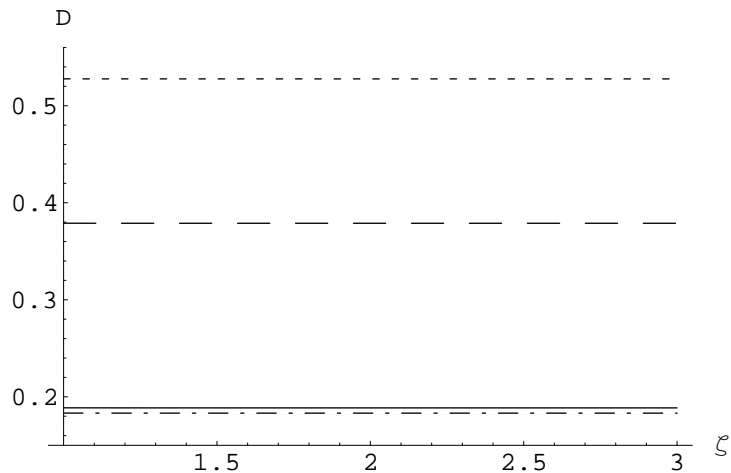}
\caption{Dispersion as function of system squeezing parameter $\zeta$,
for ten two-level atomic systems  starting in an atomic squeezed state
(Eq.   (\ref{2a.12})),    at   various   temperatures    for   a   QND
system-environment   interaction.  Here   $a=0.0$,  $\gamma_0=0.0025$,
$t=1.0$,  $j=p=5$,  $\omega=1.0$,  $\omega_c=100.0$.  The  logarithmic
dependence of $\Theta$ on  $\zeta$ (Eq. (\ref{eq:zeta})) implies a low
sensitivity of  the phase distribution to $\zeta$.   The bold, dashed,
dotted  curves   correspond  to  the  temperatures   (in  units  where
$\hbar\equiv  k_B=1$) $T=0.0$, $50.0$  and $100.0$,  respectively. The
dot-dashed curve represents unitary evolution ($\gamma_0=0$).}
\label{fig:var2}
\end{figure}

Figure  \ref{fig:var2} is analogous  to Figure  \ref{fig:var1}, except
that  the dispersion  of  $\phi$  is plotted  with  respect to  system
squeezing  parameter  $\zeta$   (Eq.  (\ref{eq:zeta})).   As  $\Theta$
appearing in the expression for $P(\phi)$ has a logarithmic dependence
on  $\zeta$ (\ref{eq:zeta}),  dispersion is  insensitive to  change in
$\zeta$ over the plotted  range.  However, as expected, the dispersion
increases  with  temperature  because   of  the  diffusive  effect  of
temperature on the phase distribution.

\begin{figure}
\includegraphics{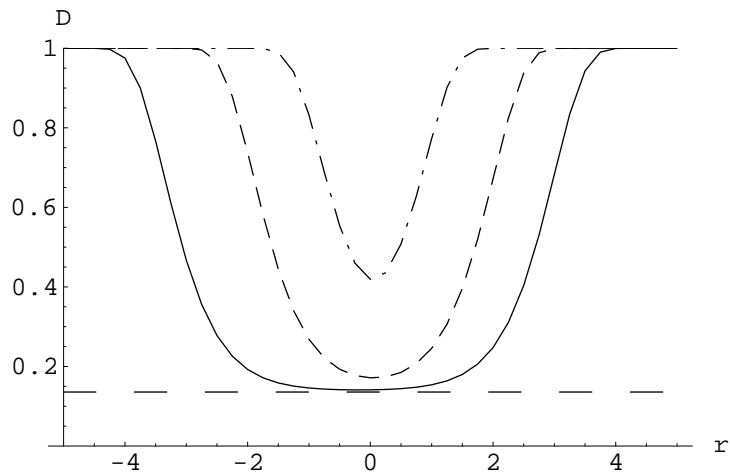}
\caption{Dispersion as  function of environmental  squeezing parameter
$r$ for  a harmonic oscillator  starting in a squeezed  coherent state
(Eq.   (\ref{eq:sqcoho}))   at   various   temperatures  for   a   QND
system-environment  interaction.    Here  $\omega=1$,  $\omega_c=100$,
$|\alpha|^2=5$,  $\gamma_0=0.0025$,  and  $t=0.1$. Here  the  parameter
$a=0$,  and   the  system  squeezing  parameters   are  $r_1=0.5$  and
$\psi=\pi/4$.  The bold, small-dashed and dot-dashed curves correspond
to temperatures  (in units where  $\hbar\equiv k_B=1$) $T=0$,  100 and
1000,  respectively.  The large-dashed  curve  corresponds to  unitary
evolution ($\gamma_0=0$).}
\label{fig:sqmodcohvar}
\end{figure}

Figure  \ref{fig:sqmodcohvar} illustrates  the behavior  of dispersion
$D$  of $\phi$  of  a  harmonic oscillator  starting  from a  squeezed
coherent  state interacting  with a  squeezed thermal  bath via  a QND
interaction,  with respect  to environmental  squeezing  parameter $r$
(Eq.  (\ref{eq:a})).   The  dispersion  is  found   to  increase  with
temperature and squeezing, tending to  the maximal value $1$. Here the
large-dashed curve,  which represents  the case of  unitary evolution,
shows no  variation with respect to change  in environmental squeezing
parameter $r$ (Eq. (\ref{eq:a})), as expected.

\begin{figure}
\includegraphics{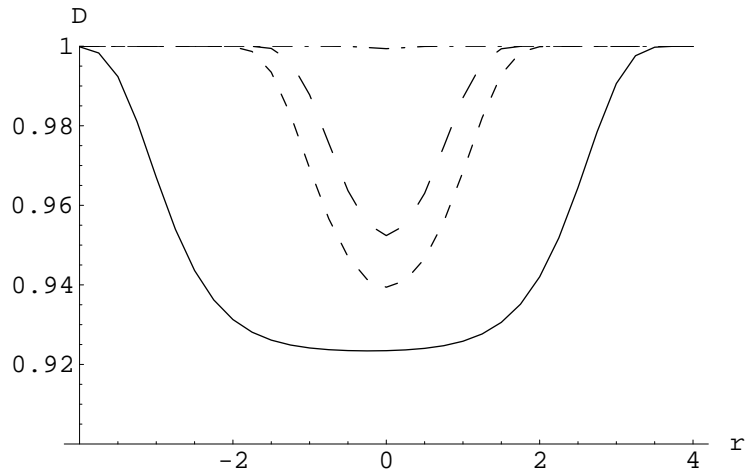}
\caption{Dispersion as  function of environmental  squeezing parameter
$r$  for a  two-level  system  starting in  an  atomic coherent  state
(Eq.  (\ref{eq:atomcohqnd})),  at   various  temperatures  for  a  QND
system-environment  interaction.    Here  $a=0.0$,  $\gamma_0=0.0025$,
$t=1.0$,           $\omega=1.0$,          $\omega_c=100$,          and
$\alpha^{\prime}=\beta^{\prime}=\pi/4$.    The   bold,   small-dashed,
large-dashed and  dot-dashed curves  correspond to temperatures  (in units
where $\hbar\equiv k_B=1$) $T=0$, 50, 100 and 1000, respectively.}
\label{fig:qndatomcohvar}
\end{figure}

Figure \ref{fig:qndatomcohvar} depicts  the behavior of dispersion $D$
of $\phi$ of  a two-level system starting in  an atomic coherent state
interacting with a  squeezed thermal bath via a  QND interaction, with
respect      to     environmental     squeezing      parameter     $r$
(Eq. (\ref{eq:a})). As  before, dispersion is found to  level out with
increase in temperature and squeezing, tending to the value $1$, which
corresponds to  a uniform  distribution. We note  that the  pattern in
this Figure is quite similar to that in Figure \ref{fig:var1}, whereas
the  use of variance  for the  data of  Figure \ref{fig:qndatomcohvar}
produces a qualitatively different pattern.

\begin{figure}
\includegraphics{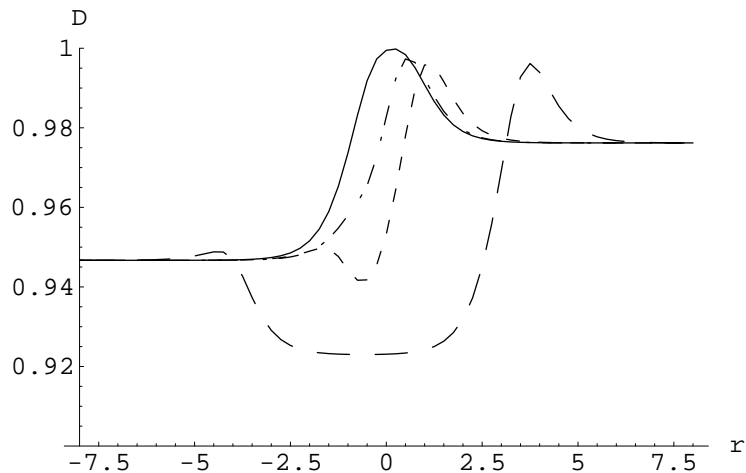}
\caption{Dispersion as  function of environmental  squeezing parameter
$r$  for a  two-level  system  starting in  an  atomic coherent  state
(Eq. (\ref{eq:atomcohqnd})), at various temperatures for a dissipative
system-environment  interaction.    Here  $\gamma_0=0.0025$,  $t=1.0$,
$\omega=1.0$,  $\omega_c=100.0$,  $\Phi=\pi/8$  [Eq. (\ref{3d})],  and
$\alpha^{\prime}=\beta^{\prime}=\pi/4$.        The       large-dashed,
small-dashed,  dot-dashed and bold  curves correspond  to temperatures
$T$  (in  units where  $\hbar\equiv  k_B=1$)  0,  100, 300  and  1000,
respectively.}
\label{fig:Disatomcohvar}
\end{figure}

Figure \ref{fig:Disatomcohvar} shows the behavior of dispersion $D$ of
$\phi$  of a  two-level system  starting in  an atomic  coherent state
interacting   with  a   squeezed  thermal   bath  via   a  dissipative
interaction,  with respect  to environmental  squeezing  parameter $r$
(Eq. (\ref{eq:a})).  While in  the case of QND system-bath interaction
(Figures        \ref{fig:var1},        \ref{fig:sqmodcohvar}       and
\ref{fig:qndatomcohvar}), the dispersion  is symmetric about $r=0$, it
is not so in this case of dissipative interaction.  Further, unlike in
the  case of  QND  interaction,  here increase  in  absolute value  of
squeezing  ($r$)  can  cause  a  decrease  in  the  dispersion.   This
illustrates the  counteractive influence of the bath  squeezing on the
thermal diffusion  of the phase distribution.   This opposing behavior
of  temperature  and squeezing  seems  to  be  generic to  dissipative
systems  \cite{bsri06}.  With  increase in  time $t$,  phase  tends to
become  randomized,  increasing  dispersion  at  any  given  squeezing
towards the maximal value of 1,  indicative of the washing away of the
non-stationary  effects due  to the  squeezed bath  \cite{bk05}.  From
Figure \ref{fig:atomCohDis}, we see  that increasing the bath exposure
time ($t$) tends to shift  and level out the distribution pattern. For
finite temperatures, the latter  effect predominates, and one observes
a steady  leveling out  with time, with  dispersion $D$ tending  to 1.
Interestingly, the use of variance in place of dispersion for the data
in Figure \ref{fig:Disatomcohvar} results in a qualitatively different
behavior.

\section{Conclusions}

In this  paper quantum phase  distributions of a number  of physically
interesting systems, interacting with their environment via a QND or a
dissipative type of coupling, are analyzed.  The system has been taken
to be either  a two-level atom (or equivalently,  a spin-$1/2$ system)
or a harmonic oscillator with the environment being modeled as a bath
of harmonic  oscillators, initially in a squeezed  thermal state, from
which  the common  thermal bath  results  may be  easily extracted  by
setting the squeezing parameters  to zero. The phase distributions are
explicitly evaluated  taking into account the effect  of the different
environmental parameters  on the dynamics of the  system starting from
various initial states.

In Section  II, we recalled  previous work on phase  distributions for
QND systems \cite{sb06} of  two-level atomic systems (Section IIA) for
different initial conditions of the  system, starting (1) in an atomic
coherent state,  and (2) in  an atomic squeezed  state; and also  of a
harmonic  oscillator  (Section   IIB)  with  the  oscillator  starting
initially in (1) a coherent  state, and (2) a squeezed coherent state.
In Section IIA, some of the above results were extended by considering
the phase distribution for multiple two-level atoms.  In particular we
studied, in Figure \ref{fig:qnd10np},  the effect of the environmental
parameters on  the distribution for  ten atoms starting in  an atomic
squeezed  state and  undergoing  a QND  system-bath interaction.   The
increase  in  bath squeezing  $r$  and  temperature  $T$ causes  phase
diffusion while the increase in the bath exposure time $t$, causes the
phase  distribution   to  diffuse  as   well  as  shift.    The  phase
distributions   are  normalized.   We   also  introduced   the  number
distribution $p(m)$, expectation of the reduced density matrix $\rho^s
(t)$  in the  Wigner-Dicke states  $|j, m  \rangle$. By  regarding the
variables $m$  and $\phi$  as the `number'  and `phase' of  the atomic
system,  the   relationship  between  the   distributions  $p(m)$  and
$P(\phi)$ may be considered as expressing complementarity in an atomic
context.

In  Section III,  the reduced  density  matrix of  a two-level  system
interacting with a squeezed thermal bath via a dissipative system-bath
interaction, resulting in a  Lindblad form of evolution, was obtained,
which reduces to the one found  by Nakazato et al. \cite{nh06} for the
case of a  thermal bath without squeezing.  This  solution was used to
study the phase distribution for the system, starting (1) in an atomic
coherent  state, and  (2)  in  an atomic  squeezed  state.  The  phase
distribution  curves preserve the  normalization of  the distribution.
The phase distribution  exhibit diffusion as well as  shift with time,
as  seen  from Figures  \ref{fig:atomCohDis}  and \ref{fig:spnp}.   An
interesting  feature   that  emerges  from   our  work  is   that  the
relationship between  squeezing and temperature effects  depend on the
type of system-bath interaction.  In the case of QND type interaction,
squeezing  and temperature  work in  tandem, and  produce  a diffusive
effect  on  the  phase  distribution.   In  contrast,  in  case  of  a
dissipative interaction, with the  reduced system dynamics governed by
a  Lindblad equation  (\ref{3a}),  squeezing tends  to counteract  the
influence of temperature, manifesting as a resistence to randomization
of  phase.  This was  noted for  example in  a comparison  between the
large- and dot-dashed curves  of Figure \ref{fig:atomCohDis}, and also
in  comparison between  the bold  and large-dashed  curves  in Figures
\ref{fig:spnp}.  A similar behavior is observed in the joint effect of
temperature and squeezing on the geometric phase of a qubit (two-level
system) interacting dissipatively  with its environment \cite{bsri06}.
Complementarity between the variables  $m$ and $\phi$, by a comparison
of  the distributions  $p(m)$ and  $P(\phi)$,  was brought  out in  an
interesting manner  for the  case of a  dissipative system-environment
interaction and seen from  a comparison of Figure \ref{fig:atomcohPfi}
with Figure  \ref{fig:atomcohpm}.  In Figure  \ref{fig:atomcohpm}, for
the case  where temperature $T=0$ and bath  squeezing parameter $r=0$,
the system tends to the pure state $|j=1/2, m=-1/2\rangle$, as seen by
the large-dashed curve.   This corresponds to the action  of a quantum
deleter  \cite{deleter}  by  means  of an  amplitude  damping  channel
\cite{nc00}.  Correspondingly the complementary distribution $P(\phi)$
is   seen   to  level   out   (the   large-dashed   curve  in   Figure
\ref{fig:atomcohPfi}), indicating complete randomization.
 
In  Section  IV,  the   quantum  phase  distribution  for  a  harmonic
oscillator in a dissipative  interaction with a squeezed thermal bath,
with  the  system starting  out  in  a  squeezed coherent  state,  was
obtained. An interesting  fact that emerged was that  under the action
of  the master equation  (\ref{4a}), which  is of  a Lindblad  kind, a
harmonic oscillator  starting in a  squeezed coherent state ends  in a
mixture that  can be expressed  as a sum  over GSCS.  A  comparison of
this  distribution  with  that  of   the  analogous  case  for  a  QND
system-bath  interaction  (Figure  \ref{fig:2plots})  brings  out  the
differing effects of the two  types of system-bath interactions on the
phase distribution.
  
In  Section   \ref{sec:appli},  as  an  application   we  studied  the
dispersion  of  phase using  the  phase  distributions conditioned  on
particular  initial  states  of  the  system.   In  the  case  of  QND
system-bath interaction,  the profile  of dispersion $D$  is symmetric
about     $r=0$,    as     seen    from     Figures    \ref{fig:var1},
\ref{fig:sqmodcohvar}  and \ref{fig:qndatomcohvar}.  In  contrast, the
profile of  dispersion is not symmetric  in the case  of a dissipative
interaction   (Figure  \ref{fig:Disatomcohvar}),  indicative   of  the
greater complexity of the latter type of interaction.  Dispersion is a
measure  of phase  fluctuations.  Since  the phase  distributions used
here  are   obtained  taking  the  effect  of   the  environment  into
consideration,  the  dispersions calculated  using  them  would set  a
realistic estimate  on phase measurements in a  number of experimental
scenarios.

We hope that the treatment  of phase distributions developed here would
be of interest both  from a technical point of view as  well as in the
context of experimental situations.

\acknowledgments

We thank Prof. Joachim Kupsch for helpful comments. We are 
also thankful to the anonymous Referee for suggestions that
have helped improve Section \ref{sec:appli}.

\end{document}